\documentclass[12pt,preprint]{aastex}
\usepackage{emulateapj5}
\usepackage{epsfig}

\newcommand{\etal}{\mbox{et al.}}
\newcommand{\ksxrb}{\mbox{KS 1731$-$260}}
\newcommand{\aqlxone}{\mbox{Aql X-1}}

\newcommand{\sixb}{\mbox{4U 1636$-$536}}
\newcommand{\sevenb}{\mbox{4U 1702$-$429}}
\newcommand{\slowb}{\mbox{4U 1728$-$34}}

\newcommand{\mxbecl}{\mbox{MXB 1659$-$298}}

\newcommand{\rxte}{{\it RXTE}}

\newcommand{\rms}{rms}

\makeatletter
\newenvironment{inlinefigure}{%
\def\@captype{figure}%
\noindent\begin{minipage}{0.999\linewidth}\begin{center}}
{\end{center}\end{minipage}\smallskip}
\makeatother

\shortauthors{Muno, \"{O}zel, \& Chakrabarty}
\shorttitle{Profiles of Burst Oscillations}

\slugcomment{To appear in \apj, v581, No.\ 1}

\begin{document}
\title{The Amplitude Evolution and Harmonic Content of Millisecond 
Oscillations in Thermonuclear X-ray Bursts}
\author{Michael P. Muno\altaffilmark{1}, Feryal
\"{O}zel\altaffilmark{2}, and Deepto Chakrabarty\altaffilmark{1,3}}

\altaffiltext{1}{Department of Physics and Center for Space Research,
       Massachusetts Institute of Technology, Cambridge, MA 02139,
       muno@space.mit.edu, deepto@space.mit.edu}
\altaffiltext{2}{School of Natural Sciences, Institute for Advanced Study,
       Princeton, NJ 08540, and Department of Physics, Harvard University, Cambridge, MA
       02138, fozel@ias.edu}
\altaffiltext{3}{Alfred P. Sloan Research Fellow}

\begin{abstract}
We present a comprehensive observational and theoretical analysis 
of the amplitudes and profiles of oscillations that occur during 
thermonuclear X-ray bursts from weakly-magnetized neutron stars in 
low mass X-ray binaries. Our sample contains 59 oscillations from six 
sources observed with the {\it Rossi X-ray Timing Explorer}. The oscillations
that we examined occurred primarily during the decaying portions of bursts, and 
lasted for several seconds each. We find that the oscillations are as large as 15\% 
during the declines of the bursts, and they appear and disappear because of to intrinsic 
variations in their fractional amplitudes. However, the 
maxima in the amplitudes are not related to the underlying flux in the burst. 
We derive folded profiles for each oscillation train to study the pulse 
morphologies. The mean rms amplitudes of the oscillations are 5\%, 
although the eclipsing source \mxbecl\ routinely produces 10\% oscillations 
in weak bursts. We also produce combined profiles from all of the 
oscillations from each source. Using these pulse profiles, we
place upper limits on the fractional amplitudes of harmonic and 
half-frequency signals of 0.3\% and 0.6\%, respectively 
(95\% confidence). These
correspond to less than 5\% of the strongest signal at integer harmonics, 
and less than 10\% of the main signal at half-integer harmonics. We then 
compare the pulse morphologies to theoretical profiles from models with 
one or two antipodal
bright regions on the surface of a rotating neutron star. We find that 
if one bright region is present on the star, it must either lie near 
the rotational pole or cover nearly half the neutron star in order to be 
consistent with the observed lack of harmonic signals. If an antipodal 
pattern is present, 
the hot regions must form very near the rotational equator. We discuss 
how these geometric constraints challenge current models for the 
production of surface brightness variations during the cooling phases of X-ray bursts.
\end{abstract}

\section{Introduction}

Flux oscillations with millisecond periods have been observed during 
thermonuclear bursts from weakly-magnetized neutron stars in nine different 
low-mass X-ray binaries (LMXBs; see Strohmayer 2001 for a review). 
The bursts occur when 
helium in the accreted material on the stellar surface begins to burn in 
an unstable regime (see Lewin, van Paradijs, \& Taam for a review). 
Therefore, it has long been 
expected that anisotropies in the burning could produce pulsations at the 
stellar spin frequency \citep[e.g.][]{sk91,bil95,str96}. 
The amplitudes of the oscillations vary between $1-50\%$ rms, with the 
largest fractional amplitudes observed in the rises of bursts when there is spectral 
evidence for growing burning regions (Strohmayer, Zhang, \& Swank 1997).
Oscillations are observed for up to 15 seconds.
Their frequencies evolve by as much as 
1.3\% during the course of a burst, usually increasing rapidly at 
first, but appearing to saturate at an asymptotic frequency before they 
disappear \citep{str97}. If we account for this 
frequency evolution, 70\% of the oscillations appear coherent \citep{mgc02}, 
and the asymptotic frequencies are 
stable to a few parts in a thousand in bursts separated by several years
\citep{sm99,mun00,gil02}. Although the underlying clock may not be 
perfectly stable, it is nonetheless remarkably good \citep{mgc02,sm02}. 
This strongly suggests that
the oscillations are produced by patterns in the surface brightness of these 
rotating neutron stars. 

Several mechanisms have been proposed to explain the frequency evolution of 
the burst oscillations. \citet{str97} suggested that the 
oscillations originate from hot regions on a burning layer that 
expands and decouples from the neutron star when the nuclear burning 
commences. The oscillations are observed when the burning layer begins to 
cool and contract, causing them to increase in frequency as the layer 
re-couples to the neutron star. However, calculations suggest that
a rigidly rotating, hydrostatically expanding burning layer produces too 
small a frequency drift \citep[see][]{cb00,cum02}. Recently, Spitkovsky, Levin, \& 
Ushomirsky (2002) pointed out 
that vortices could form in a geostrophic flow moving against the rotation
of the neutron star, driven by the combination of the Coriolis force
and a pressure gradient between the equator and the poles. Like their
counterparts on Earth and on Jupiter, such vortices could conceivably
appear as light or dark regions on the neutron star. The frequency drift in
this model is attributed to the slowing of the geostrophic flow
as the burning layer cools. Finally, Heyl (2002) has proposed that
global oscillation modes could propagate as waves on the neutron star
ocean. The velocity with which these modes travel is extremely
sensitive to the vertical density and temperature structure, as well as
to the surface composition, and could thus change during the
burst. Indeed, in all these models, the mechanisms producing the
pulsations depend strongly on the properties of the neutron stars, and
the oscillations offer a powerful probe of the physical conditions 
in their outer layers. In particular, the pulsations may reveal the stellar 
spin frequency and surface gravity.

All of the above models assume that the oscillations occur near the 
rotational frequency of the neutron star ($\nu_{\rm spin}$). If this
is the case, then the burst oscillations trace the history of accretion 
torques on the neutron stars in these LMXBs, which are thought to be the
progenitors of recycled millisecond radio pulsars \citep{alp82,rs82}. 
The frequencies of the oscillations ($\nu_{\rm burst}$) are distributed 
evenly between 270--620~Hz 
\citep{mun01}. However, there are observational and theoretical reasons to 
suspect that the subset of the oscillations with 
$\nu_{\rm burst} \approx 500-600$~Hz occur at twice the spin 
frequency, corresponding to two antipodal bright regions on the neutron star 
(Strohmayer et al. 1996; Miller, Lamb, \& Psaltis 1998; Miller 1999;
van der Klis 2000). 
%However, two observational results have raised the possibility 
%that some of the oscillations occur at twice the spin frequency of the
%neutron star. First, there appears to be a coincidence between the 
%frequencies of the burst oscillations ($\nu_{\rm burst}$) and 
%the difference frequency of pairs of kilohHertz quasi-periodic 
%oscillations (kHz QPOs, $\Delta\nu_{\rm kHz}$) observed in the persistent 
%emission from 18 LMXBs, such that 
%$m\times\Delta\nu_{\rm kHz} \approx \nu_{\rm burst}$ 
%(see van der Klis 2000 for a review). %\citep[see][]{vdk00,gal00}. 
%The difference frequency, $\Delta\nu_{\rm kHz}$, is always near 300~Hz, 
%with $m=1$ when $\nu_{\rm burst} \approx 300$~Hz, and $m=2$ when 
%$\nu_{\rm burst} \approx 600$~Hz \citep[but see][]{vdk00,gal01}. 
%It has been suggested that 
%$\Delta\nu_{\rm kHz} \approx \nu_{\rm spin}$ (Strohmayer et al. 1996, 
%Miller, Lamb, \& Psaltis 1998).
%Second, a signal at $0.5\times\nu_{\rm burst}$ was reported in a set of
%bursts from 4U~1636$-$536 \citep{mil99}, a source with 
%$\nu_{\rm  burst} \approx 2\times\Delta\nu_{\rm kHz}$. 
%However, this signal has not
%been found in independent sets of bursts from this source \citep{str01}. 
Determining whether $\nu_{\rm burst} = 2\times\nu_{\rm spin}$ in 
the subset of sources with 600~Hz oscillations is particularly important, 
because if 
%$\Delta\nu_{\rm kHz} \approx 
$\nu_{\rm spin} \approx 300$~Hz in all
nine of these LMXBs, then some mechanism may be limiting the maximum
spin frequencies of the neutron stars \citep[e.g.,][]{wz97, bil98}.

The propagation of photons from the surface of a rapidly rotating
neutron star is affected by general relativity, 
time delays, and Doppler shifts, and hence these oscillations carry
signatures of the physical parameters of the neutron star to an observer
(Miller \& Lamb 1998; Braje, Romani \& Rauch 2000; Weinberg, Miller, \& 
Lamb 2001; Nath, Strohmayer, \& Swank 2002). In particular, strong 
gravitational lensing by the neutron star allows a bright region on its 
surface to be seen for a large fraction of the rotational period. 
This, in general, suppresses the amplitudes and reduces the harmonic content 
of any resulting oscillations 
\citep[see][]{wml01,oze02}. On the other hand, Doppler and time delay effects 
cause the pulse profiles to be more asymmetric and narrowly-peaked, 
which increases the amplitudes and the harmonic content of the 
oscillations \citep{wml01,brr00}. Thus, the properties of the burst 
oscillations can constrain the emission geometry 
and the compactness of the neutron star.

In this paper, we present a comprehensive observational and 
theoretical investigation of the amplitudes and profiles of burst 
oscillations, seeking to constrain the geometry of the emission. 
We focus on oscillations observed during the peak and decline of X-ray 
bursts with the {\it
Rossi X-ray Timing Explorer} \rxte. These signals
are present for tens of seconds, and thus provide excellent statistics 
to constrain the pulse profiles and amplitudes of the pulsations 
\citep[see][]{nss02}. 

In Section~\ref{sec:obs}, we examine the amplitude evolution and the 
profiles of the oscillations. In Section~\ref{sec:mod}, we present
theoretical predictions of the signal that an observer would see from
one or two bright regions on the surface of a rapidly rotating neutron
star. We explore a range of parameters relevant to the burst oscillations,
and explicitly take into account the response of the \rxte\
detectors to allow a direct comparison with the data 
\citep[compare][]{ml98,brr00,wml01}. In
Section~\ref{sec:res}, we place constraints on the location and size
of bright regions on the neutron star surface by comparing the
theoretical calculations to observations. Finally, we discuss the
implications of these constraints for the various models proposed to
explain the burst oscillations.

\section{Observations\label{sec:obs}}

Our analysis used observations with the Proportional Counter Array
(PCA; Jahoda et al. 1996) on \rxte.  The 
PCA consists of five identical gas-filled proportional
counter units with a total effective area of 6000~cm$^2$ and
sensitivity in the 2.5--60~keV range.  The detector is capable of
recording photons with microsecond time resolution and 256-channel
energy resolution.   The data were recorded
in a wide variety of data modes with different time and energy
resolutions, depending upon the details of the original proposed
programs and the available telemetry bandwidth.   For all of the analysis
presented here, we converted the photon arrival times at the
spacecraft to Barycentric Dynamical Time (TDB)
at the solar system
barycenter, using the Jet Propulsion Laboratory DE-200 solar system 
ephemeris (Standish et al. 1992). 

We searched the entire {\em RXTE} public data archive for
X-ray bursts from 8 neutron stars\footnote{Oscillations in bursts
associated with MXB~1743$-$29 have been observed during observations
of the bursting pulsar GRO~J1744$-$28.  A search for these bursts was
not part of the analysis presented here.} that are known to exhibit
burst oscillations (see Table 1 and Muno et al. 2001).  As of
September 2001, we have identified a total of 159 X-ray bursts from
these 8 sources.  Each of these bursts was then searched for millisecond
oscillations as described in \citet{mgc02}. Out of the 68 pulse trains detected, 
59 persisted continuously for $> 2$~s and therefore warranted a more detailed analysis 
(Table~\ref{tab:harm}). The majority of these continuous pulse trains were observed 
during the declining portions of the bursts. Although 39 of the 68 oscillation 
trains were detected during the rises of bursts, only 17 lasted continuously through 
to the tails. Six oscillations appeared only in the rises of bursts and persisted 
for less than 1~s, precluding further analysis. The remaining 16 oscillations were
observed during the rises and tails of bursts, but disappeared during the peaks. We 
were only able to analyze these in the tails of the bursts when they re-appeared.

To examine the amplitudes and profiles of the oscillations, 
we used data containing a single energy channel (2.5--60 keV) and 
$2^{-13}$~s (122 $\mu$s) time resolution. 
We modeled the frequency evolution of the oscillations via 
a phase connection method commonly used in pulsar studies 
\citep{mt77}.  In this technique, we fold the data in short 
intervals (0.25--0.5 s) about a trial phase model, which is then refined 
through a least squares fit to the residuals. This provides excellent 
frequency resolution, 
\begin{figure*}[t]
%\epsscale{0.8}
%\plotone{f1.eps}
\centerline{\epsfig{file=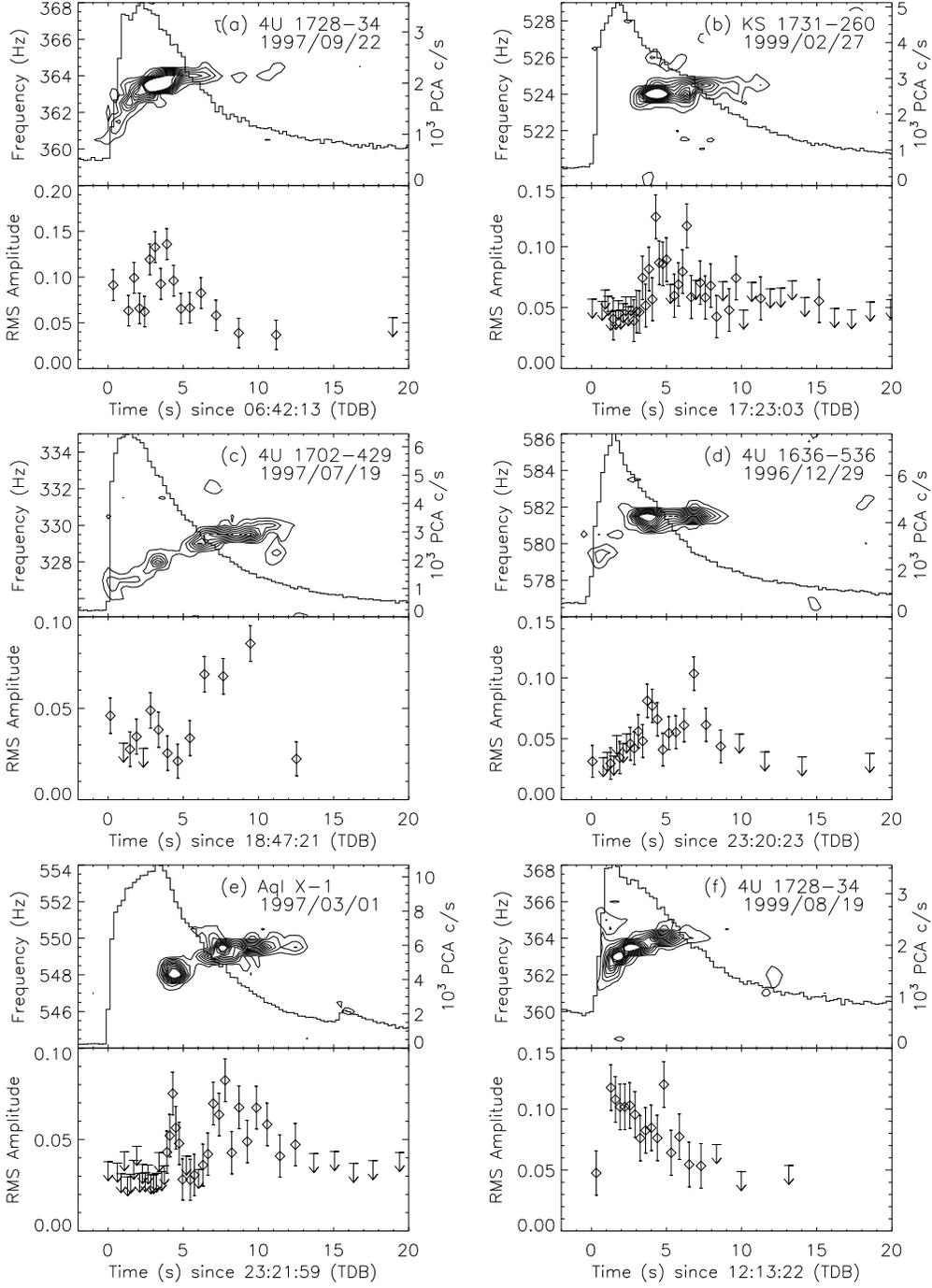,width=0.78\linewidth}}
\caption{
Six examples of oscillations during X-ray bursts. 
{\it Top Panels}: Contours represent the Fourier power as a function of time 
and frequency, computed from the power spectra of 2 s intervals
of data every 0.25 s throughout the burst. A Welch function was used to 
taper the data to reduce sidebands in the power spectrum due to its finite 
length \citep{pre92}. The contour levels are at powers of $0.02$ in 
single-trial probability starting at a chance occurrence of $0.02$.
The PCA count rate (2--60~keV) 
is also plotted, referenced to the right axis. {\it Bottom Panels}: The 
fractional rms amplitude of the oscillations as a function of time 
during the burst. The data were folded in intervals that contained a 
constant count rate, such that each interval is sensitive to oscillations
of a constant fractional amplitude (between 4--10\%).
The amplitudes of the oscillations do not 
appear to be correlated with the amount of flux from the underlying burst.
}
\label{fig:ampev}
\end{figure*}
and a statistical measure of how well the model 
reproduces the data \cite[see][for a further description]{mgc02}.
We folded the 59 continuous oscillation trains about the best-fit frequency models of 
\citet{mgc02}. 

We measured the amplitudes of the oscillations and their harmonics by 
computing a Fourier power spectrum of the folded profiles. If we normalize 
the power according to \citet{lea83}, the fractional rms amplitude at any
multiple of the oscillation frequency is
\begin{equation}
A_n = 
\left({{P_n}\over{I_\gamma}} \right)^{1/2}{{I_\gamma}\over{I_\gamma-B_\gamma}},
\label{eq:acont}
\end{equation}
where $P_n$ is the power at the $n$th bin of the Fourier spectrum, 
$I_{\gamma}$ is the total number of counts in the profile, and $B_\gamma$
is the estimated number of background counts in the profile.
This equation is valid so long as the phase and frequency of the 
oscillation is known, as it is by design for our folded profiles.
Uncertainties and upper limits on the amplitudes are computed taking into
account the distribution of powers from Poisson noise in the spectrum, 
using the algorithm in the Appendix of \citet{vau94}.

The models used were low-order polynomials or exponential functions that 
reproduced the phase evolution well in 70\% of the oscillations.
In the remainder of the oscillation trains, there was evidence both for 
discrete phase jumps and for phase evolution that was only piecewise smooth.
Additional power in principle could be recovered from these oscillations with 
more complicated frequency models, but we believe that the advantage would 
be small.
For instance, we compared the powers we detected at the fundamental frequency 
to those obtained by \citet{sm99} for two oscillations that are common 
to both samples (bursts on 1997 July 26 and 30 from \sevenb). 
\citet{sm99} used a $Z^2$ technique to determine the 
exponential frequency models that recovered the most power in the 
oscillations, while our technique determined that polynomial models best 
reproduced the phases of the oscillations. Nonetheless, we find that the 
values for the amplitudes of the oscillations are consistent within 1\% 
fractional \rms\ \citep[see Figures 1 and 2 in][]{sm99}. This is 
similar to the uncertainty in the oscillation power that is introduced
by Poisson counting noise \citep{vau94}, indicating that
slight modifications to the frequency models introduce only minor differences
in the amplitudes derived for the oscillations.

We also used data with at least 16 energy channels between 2--60~keV
in order to produce spectra for 
0.25 s intervals during each observation. We estimated the 
background to the burst emission to be the average persistent flux during 
the 16 s prior to the burst \citep[e.g.,][]{kul02}.
The detector response was estimated using PCARSP in FTOOLS version 
5.1.\footnote{see http://heasarc.gsfc.nasa.gov/lheasoft/}
Each spectrum was fit with a blackbody modified at low energies
by a constant interstellar absorption. The color temperatures ($T_{\rm col}$)
from these spectral fits were used when computing theoretical 
lightcurves (Section~\ref{sec:mod}).

\subsection{Amplitude Evolution of the Oscillations}

In order to examine how the amplitudes of the oscillations evolve as a 
function of time during each burst, we folded short intervals of data
using the frequency models of \citet{mgc02}. Each interval
contained a constant total number of counts from the burst emission, and
therefore had
a constant sensitivity to oscillations of a given fractional amplitude. 
In order to place upper limits on the amplitudes in those intervals when 
oscillations were not detected or were not part of the continuous portion of the
train, we assumed that the 
frequency remained constant at the value we derived for the detection nearest 
in time.
We note that the technique we have used is most effective for analyzing long 
oscillation trains during the decaying portions of bursts. We are unable to 
measure the very large amplitudes 
that are sometimes observed during the rises of bursts \citep[e.g.][]{szs97}.
The small count rates during these periods required us to use longer 
time bins from which we only can measure the average amplitude of a rapidly declining 
signal. The amplitude evolution of oscillations during the rises of bursts have
been studied by, for example, \citet{szs97} and \citet{nss02}.

Of the 59 oscillation trains with long ($> 1$~s) continuous portions, 34 exhibited 
significant variations in their amplitudes that could be measured when
folding the data in bins with constant count rate. The amplitudes of 
the rest of the oscillations did not vary significantly, and always remained 
just above the detection threshold (4--10\% rms).
Figure~\ref{fig:ampev} displays six examples of oscillations (top panels)
and their associated amplitude evolution (bottom panels) that are 
representative of our sample as a whole.
In most cases, local maxima in the fractional amplitudes occur as the flux 
from the burst decays. Figures~\ref{fig:ampev}a and b illustrate two bursts in
which the largest amplitudes
are observed 1--2 s after the flux from the burst begins to decline. 
This behavior is observed in 13 of the oscillation trains from our sample. 
In Figure~\ref{fig:ampev}c the largest fractional amplitude is observed
much later in the burst, 8 s after the flux from the burst starts to decay.
In 10 of the oscillations, peaks in the amplitudes are observed several
seconds into the decays of the bursts. The oscillations in 
Figures~\ref{fig:ampev}d and \ref{fig:ampev}e 
exhibit two separate and significant maxima in their amplitudes during
the burst decays. In total, 3 oscillations exhibit this behavior;
less significant secondary maxima may also be present in 
Figures~\ref{fig:ampev}b and \ref{fig:ampev}f.

Strong oscillations are also observed in the peaks of X-ray bursts, 
although less often. In Figure~\ref{fig:ampev}f, the amplitude of the 
oscillation is largest when the flux from the burst is highest, and declines 
steadily as the burst decays. Maxima are observed during the peaks 
of bursts in only 5 cases.  This behavior is to be distinguished from 
instances where the amplitude declines {\it as the burst flux is rising} 
\citep[e.g.,][]{szs97}. 

In all of the examples in Figure~\ref{fig:ampev}, the fractional amplitudes of 
the oscillations drop suddenly below our detection threshold by 15 s into the 
burst. Their disappearance is due to a genuine decrease in their
amplitudes, as opposed to a lack of sensitivity when the flux from a burst
is low. Besides this general trend, the amplitudes of 
the oscillations are not correlated with the underlying flux from the bursts. 

We also measured the temperature, $T_{\rm col}$, of each burst as a function
of time, and interpolated it 
onto the times of each interval with a constant total number of 
counts. The amplitudes of the oscillations are not correlated with the 
temperatures of the burst emission (not shown).
The majority of oscillations appear when
the burst emission has color temperatures of $T_{\rm col} \sim 2-3$~keV.
We shall use these as fiducial values
when we simulate the emission from a hot region on a neutron star
in Section~3.

\subsection{The Profiles of the Burst Oscillations}

We examined the profiles of each oscillation train by folding 
the data about the best-fit phase model of \citet{mgc02}.
To allow for the possibility that the strongest signal is observed
at $2\times\nu_{\rm spin}$, in practice we  
folded the data with a frequency one-half that predicted by the
phase 
\begin{figure*}[t]
%\epsscale{1.0}
%\plotone{f2.eps}
\centerline{\epsfig{file=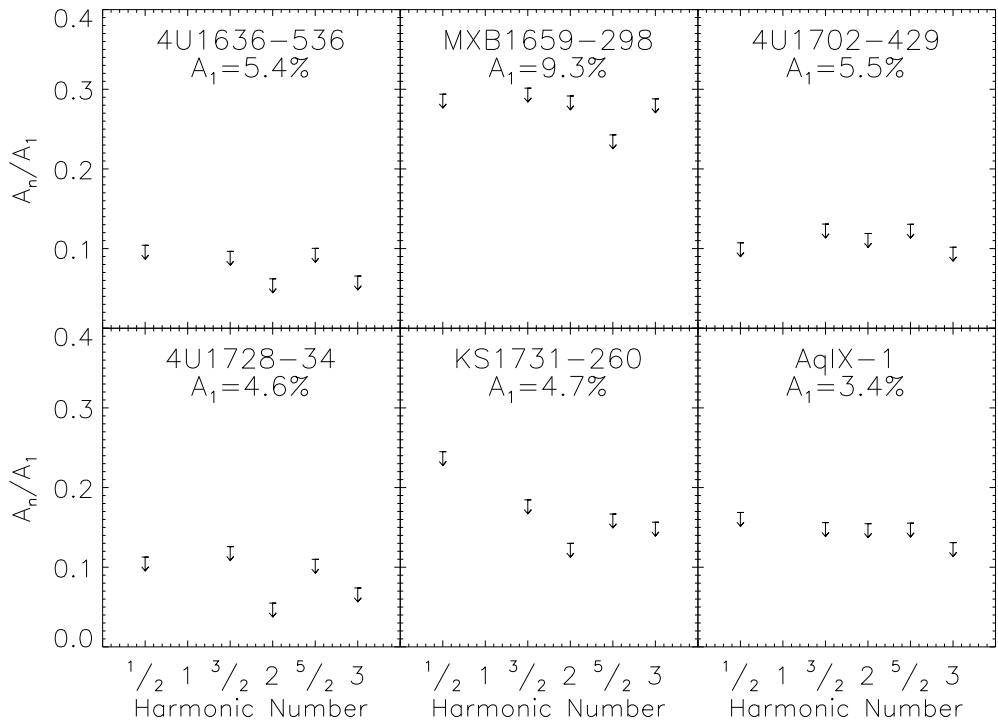,width=0.8\linewidth}}
\caption{
Upper limits (95\% confidence) on the fractional amplitude of signals 
present at integer and half-integer multiples of the strongest signal ($A_1$), 
from combined pulse profiles using all of the oscillations in each source.
}
\label{fig:ulharm}
\end{figure*}
model. A typical oscillation train contains 68,000 photons.
Since we modeled the phase evolution of each burst directly, 
the relative phases of each oscillation train were known, so we also 
produced summed profiles for each source. 
In searching for half-frequency signals, there is an 
uncertainty of one-half cycle between oscillations from different bursts, 
so we coherently added $n$ oscillations in $2^{n-1}$ combinations to 
account for our lack of knowledge of the true phase. For
\sixb\ and \slowb, we only summed those profiles with more than 
$5\times10^4$ total counts. This reduces the number of trials we needed
to search while only slightly reducing the signal-to-noise in the profile,
so that we obtain the strictest limits on the harmonic content.

We then computed the Fourier 
transform of ({\it i}) each of the 59 profiles individually and 
({\it ii}) the sum of all profiles from each source, and searched for 
possible signals at 0.5, 1.0, 1.5, 2.0, 2.5 or 3.0 times the main 
oscillation frequencies. 
We considered a signal that had less than a 32\% chance of occurring randomly 
in our entire search as a detection.

We find that the typical oscillation train has an \rms\ amplitude of 5\%. 
The most significant
oscillation has a power (normalized according to the criteria of
Leahy et al.~1983) of 615 in 100,000 
counts from a burst from \sixb\ on 2000 June 15, which translates to an 
8\% oscillation amplitude.
Oscillations from the eclipsing source \mxbecl\ have 
amplitudes of 10\%, although the bursts are faint (10,000 photons 
in a typical folded profile).
The amplitudes from the summed profiles are listed in Table~\ref{tab:harm}. 
The most significant signals
are from \sixb\ and \slowb, since we modeled 17 oscillation trains from 
\sixb\ and 24 from \slowb. The summed waveform from \sixb\ contains
$1 \times 10^6$ photons and has a Leahy-normalized power of 2580. 
The profile from \slowb\ contains $1.5 \times 10^6$ 
photons with a power of 2430.

We find no evidence for signals at half-integer or integer multiples of 
the oscillation frequencies.\footnote{In particular, we do not detect a 
signal at $A_{1/2}$ from \sixb. \citet{mil99} has previously reported
the detection of such a signal in combined profiles from the first 0.75 s 
of a particular subset of bursts from this source \citep[see also][]{str01}.}
Typical individual oscillations provide upper 
limits on the fractional \rms\ amplitudes of harmonic signals ($A_n$) of 1.6\% 
(95\% confidence), but they can be as low as 0.7\% in the brightest bursts 
from \aqlxone. These values are consistent with those reported by 
\citet{sm99} and \citet{mun00}. 
Upper limits from the summed profiles range from 2.5\% in
\mxbecl, for which we examined oscillations in only three weak bursts, 
to 0.3--0.6\% for \sixb\ and \slowb, for which we combined 
oscillations from many bright bursts (Table~\ref{tab:harm}). 
In Figure~\ref{fig:ulharm} we display the ratio of the upper limits on 
the amplitudes $A_n$ to the amplitude of the largest signal $A_1$. 
The strongest constraints are again obtained for \slowb\ and \sixb,
for which any integer harmonic signal must be less than 5\% of the 
amplitude of the detected signal, and any half-integer multiple of the
main frequency must be less than 10\% of the amplitude of the main signal. 

We took care to establish that the harmonic content would not be reduced 
when producing a mean profile. We 
\begin{inlinefigure}
\centerline{\epsfig{file=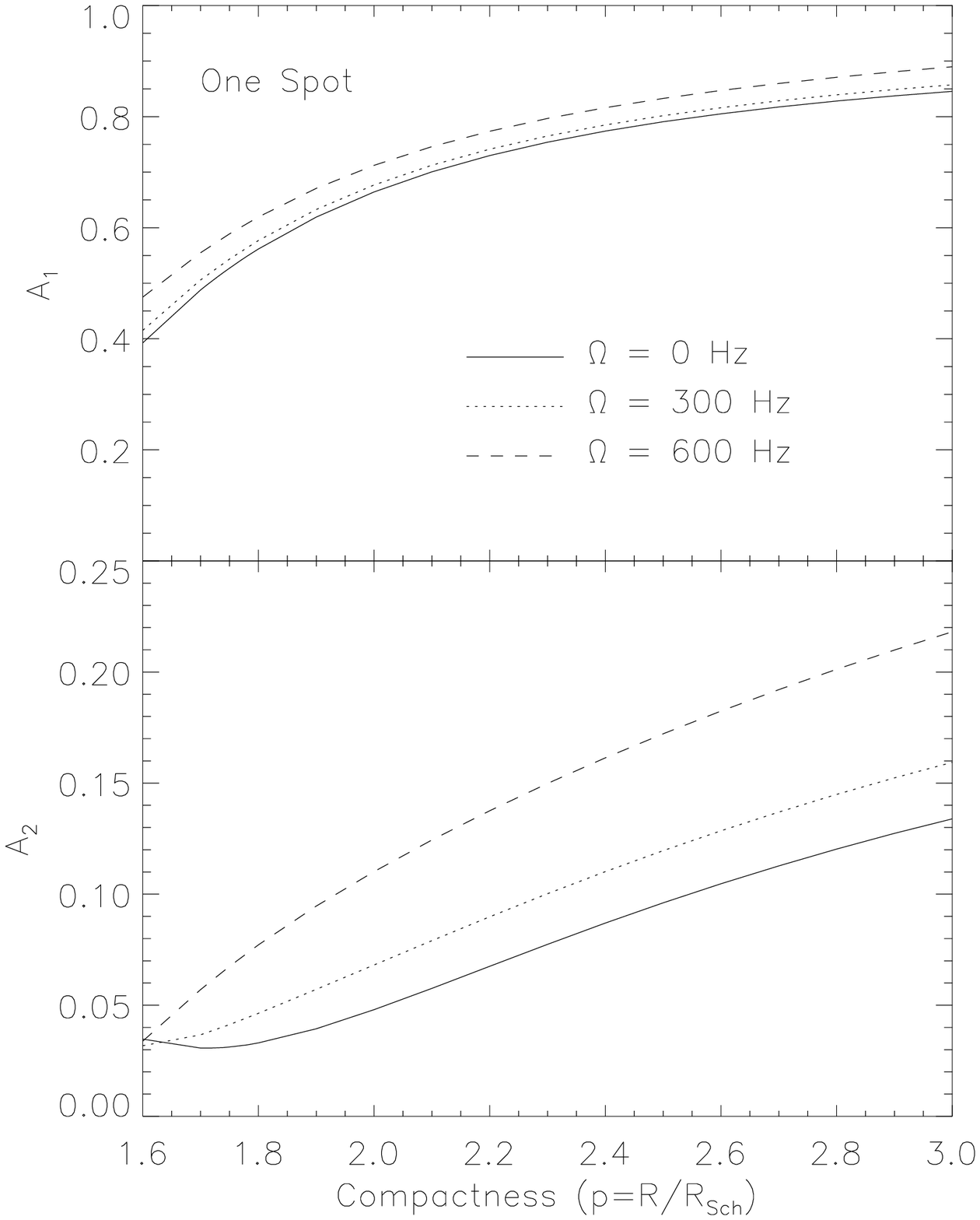,width=0.95\linewidth}}
\caption{
The predicted amplitudes of the oscillations as a function of the
compactness $p$ of the neutron star, for a few values of the spin
frequency $\Omega$. One hot region of size $\rho=60^\circ$ and of uniform 
temperature $T_{\rm col} = 3.0$~keV (at the neutron star) is 
located on the rotational 
equator ($\alpha = 90^\circ$), with the observer also viewing along the 
equator ($\beta = 90^\circ$).
Increasing the radius (higher $p$) or the spin frequency ($\Omega$) 
of the star generally increases the amplitudes $A_1$ and $A_2$. The amplitudes
$A_1$ from two bright regions at $\alpha = \beta = 90^\circ$ will be the
same as $A_2$ in this figure.
}
\label{fig:compact}
\end{inlinefigure}

\noindent
searched power 
spectra of short ($< 2$ s) intervals of data, and did not find evidence for 
signals at any multiples of the oscillation frequencies with amplitudes 
greater than 5--15\%. We also examined the folded profiles of the 
short intervals of data used in Section~2.1, and they are all consistent 
with sinusoidal signals to 5--10\%. We concluded that there are 
no gross changes in the pulse profiles with time. From the theoretical
standpoint, we used the simulations described in Section~3 to confirm that the 
relative phases of the fundamental and harmonic signals do not vary as we 
change the parameters of a bright region on the neutron star surface 
(such as its location, temperature, and size).
Although the amplitudes of harmonic signals may change, they
can not add destructively, except in the special case where the region 
covers on average half of the neutron star. Therefore,
the average profiles we measured accurately reflect the mean harmonic 
content of the oscillations, if 
they originate from a brightness asymmetry on a rotating neutron star.

\section{Models\label{sec:mod}}

In order to interpret our observational results and to place
constraints on theoretical models of burst oscillations, we calculated
lightcurves from an anisotropic temperature distribution on the
surface of a rapidly rotating neutron star using the techniques
outlined by Pechenick, Ftaclas, \& Cohen (1983), \citet{brr00}, 
and \citet{wml01}. 
We considered single and antipodal circular regions to describe the temperature patterns 
on the stellar surface. We denoted the angular radius of the circular hot
region(s) by $\rho$, and its (their) location by an angle $\alpha$
from the rotational axis of the neutron star. The angle between the 
line-of-sight of the observer and the spin axis of the neutron star was 
defined as $\beta$.

We assumed that the bright regions emit as blackbodies of a uniform 
temperature $T_{\rm col}$, and that the rest of the neutron star is dark. The 
latter choice will produce the largest amplitude
oscillations; we confirmed that assuming the rest of the neutron star
is warm decreases the amplitude of oscillations, but does not change the 
relative amplitude of the harmonics. The choice of a blackbody spectrum 
is reasonable for our
purposes because the burst spectra can be adequately modeled as such in 
the PCA bandpass. In reality, however, both the spectra and the
angular distribution of surface emission are affected by the neutron
star atmosphere (e.g., London, Taam, \& Howard 1984; Madej 1991). 
To roughly account for these effects, we adopted the Hopf
function \citep[][pp. 76--79]{cha60} to describe the beaming of radiation 
emerging from the
surface, because the atmosphere is scattering-dominated during a burst.

Photons were propagated from the neutron star to the observer through a
Schwarzschild metric for an object with a compactness $p = Rc^2/2GM$. 
Time delay and Doppler effects were computed assuming a neutron
star spin frequency of $\Omega$. Although the Schwarzschild metric is
not strictly correct for the spacetime exterior to a rotating 
neutron star, we
utilized it because the exact metric for a neutron star 
depends on its unknown structure. The correction introduced by
considering a Kerr metric instead, for example, is of order of a few
percent \citep{brr00}. We neglected this effect in our calculations,
since both the corrections from a metric appropriate for a rapidly-rotating 
neutron star and the measurement
uncertainties in the observed waveforms are of the same magnitude.

For each set of parameters, we produced light curves in 40 phase bins,
and in 64 energy bins logarithmically spaced between 0.01--25~keV. The
observed spectrum is simply a sum of blackbodies whose temperatures
are multiplied by Doppler factors; therefore, the signal from a bright region
of arbitrary temperature could be obtained by rescaling from a calculation 
with $T=1$~keV. For each phase, the resulting spectra were folded through
a fiducial PCA response matrix, which we generated for Proportional Counter
Unit 2 during
December 1999 (gain epoch 3), in order to obtain predicted lightcurves
that can be compared directly to the observations.  To analyze these
theoretical PCA lightcurves, we used the same Fourier technique that we 
applied to the observational data to measure the amplitude 
of the oscillation and its harmonics (Section~\ref{sec:obs}).

\subsection{The Amplitude of the Oscillations}

In total, our simulations included eight parameters that can affect
the amplitudes of the oscillations and their harmonics: 
the compactness and the spin frequency of the neutron star; the number, size,
position, and temperature of the hot regions; the angular distribution of
the emission from the hot region; and the viewing angle
of the observer. 
\begin{figure*}[t]
%\epsscale{1.0}
%\plotone{f4.eps}
\centerline{\epsfig{file=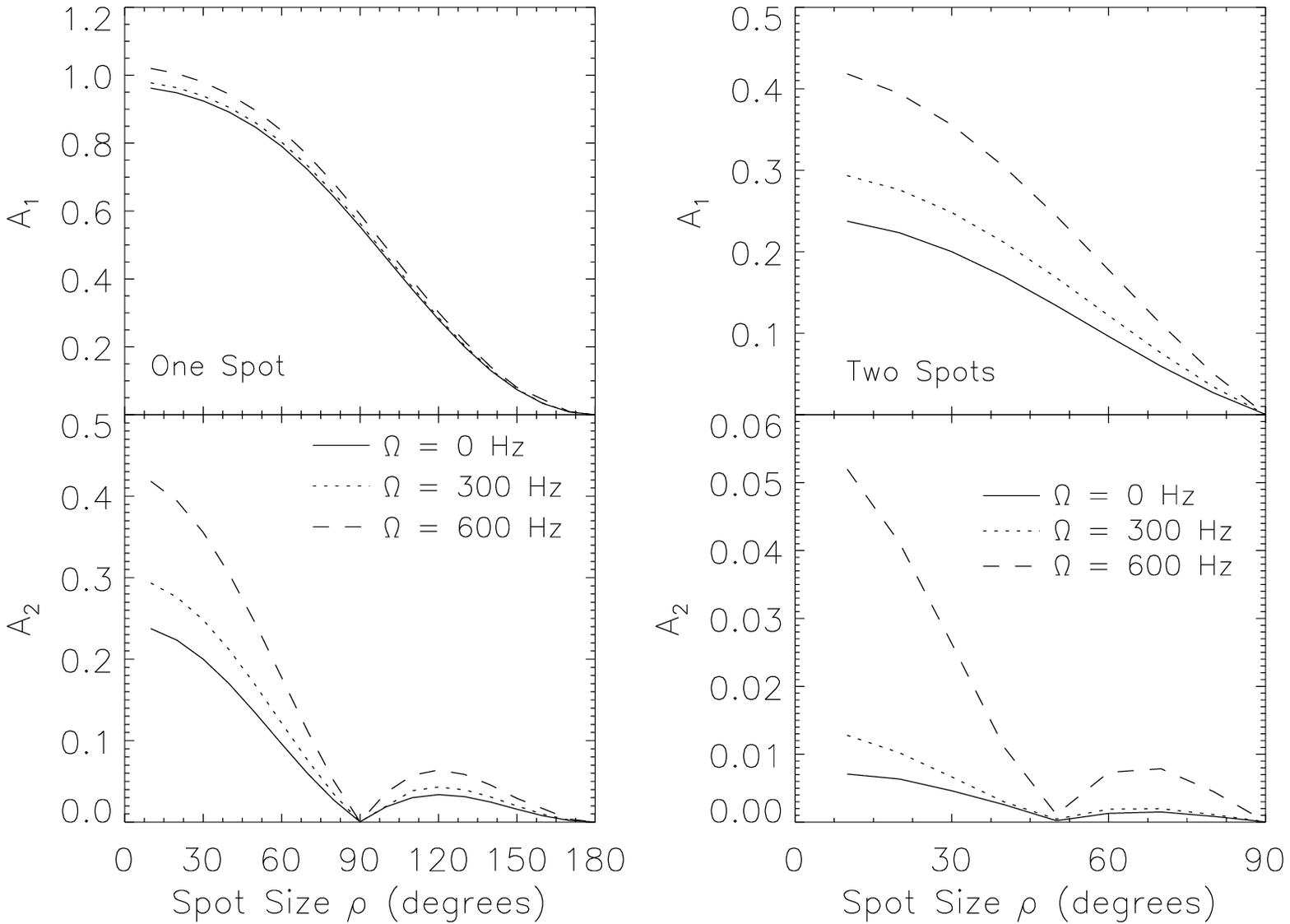,width=0.8\linewidth}}
\caption{
The predicted amplitudes of oscillations and their harmonics from 
one or two circular bright regions, with temperature $T_{\rm col}=2.3$~keV
(at the observer) as a function 
of their of size $\rho$ and of the spin frequency $\Omega$. 
Here, $\alpha = \beta = 90^\circ$.
{\it Left Panels}: Amplitudes from one bright region, where $A_1$
occurs at the spin frequency.
{\it Right Panels}: Amplitudes from two bright regions, where 
no signal appears at the spin frequency, and the strongest signal $A_1$ 
occurs at twice the spin frequency. 
In all cases the amplitudes
of the oscillations increase with larger $\Omega$, and generally decrease 
with increasing $\rho$.
}
\label{fig:rho}
\end{figure*}
We will focus on  the compactness of the star,
the geometric parameters of the hot spot, and the observer's line of sight in
this section. We will briefly summarize the results for the other parameters.
Similar studies have been carried out previously 
\citep{ml98,wml01,oze02}. In the present work, we also take into account 
the response of the PCA, which allows us to compare the theoretical
light curves directly to the properties of the oscillations. Including
the PCA response increases the amplitude of
the oscillations and their harmonics by a few percent \citep[compare][]{ml98}.

In Figure~\ref{fig:compact} we illustrate the effects of varying the 
compactness of the star $p$ for a few values of the
neutron star spin frequency, $\Omega$. We have assumed that there is a
single bright region with a temperature $T_{\rm col} = 3.0$~keV 
(at the surface), size $\rho = 60^\circ$, is located at the
equator ($\alpha = 90^\circ$), and is viewed along the equator 
($\beta = 90^\circ$).
As the compactness of the neutron star increases,
the amplitude of the strongest signal $A_1$ decreases monotonically, 
because the curved photon trajectories allow the observer to see the 
bright region even when it is behind the limb
of the neutron star. However, the amplitude of the harmonic 
signal $A_2$ reaches a minimum at $p=1.7$, and increases for smaller 
values of $p$ because the neutron star lenses the bright region when it 
is on the opposite side of the star from the observer \citep{pfc83, ml98}. 
The signal at $2\times\nu_{\rm spin}$ from two antipodal bright regions 
with $\rho=60^\circ$ located 
at $\alpha=\beta=90^\circ$ is identical to $A_2$ in Figure~\ref{fig:compact} 
\citep[see below and][]{wml01}.

In all cases, increasing the spin frequency
increases the amplitude of the harmonic more than that of the fundamental.
Moreover, the minimum in the amplitude of the harmonic at $p=1.7$ 
disappears when the neutron star is spinning rapidly. This is because 
the light travel time of the photons becomes comparable to the
rotational period of the star, which delays the arrival of the harmonic 
peak formed by strong lensing. Thus, the power in $A_2$ is transferred to  
higher harmonics for a rapidly spinning neutron star.

We adopted a fiducial value of $p=2.5$ for the rest of the calculations,
corresponding to a neutron star of mass 1.4 $M_\odot$ and radius 
10~km. If instead we choose $p=2.0$, the amplitudes of the fundamental signals
decrease by a factor of 1.2 in the following figures, and the amplitudes of 
the harmonic signals decrease by a factor of 1.5--2.0 
(Figure~\ref{fig:compact}). This decrease in the amplitude would not change
qualitatively the results that we describe in Section~\ref{sec:res}.

We then explored the effects of changing the temperature and 
angular dependence of the emission from the hot region, in order to 
select fiducial values for the rest of our study.
\citet{ml98} have shown that varying the temperature of the emitting
region only changes the amplitudes of oscillations when the peak of the 
X-ray spectrum lies at energies lower than the response of the detector. 
For the temperature ranges of the observed bursts, 
$T_{\rm col} \sim 2-3$~keV, we find that the amplitudes are 
essentially constant as a function of $T_{\rm col}$, since 
most of the bolometric flux is emitted in the 2--60~keV PCA bandpass. 
\begin{figure*}[t]
%\plotone{f5.eps}
\centerline{\epsfig{file=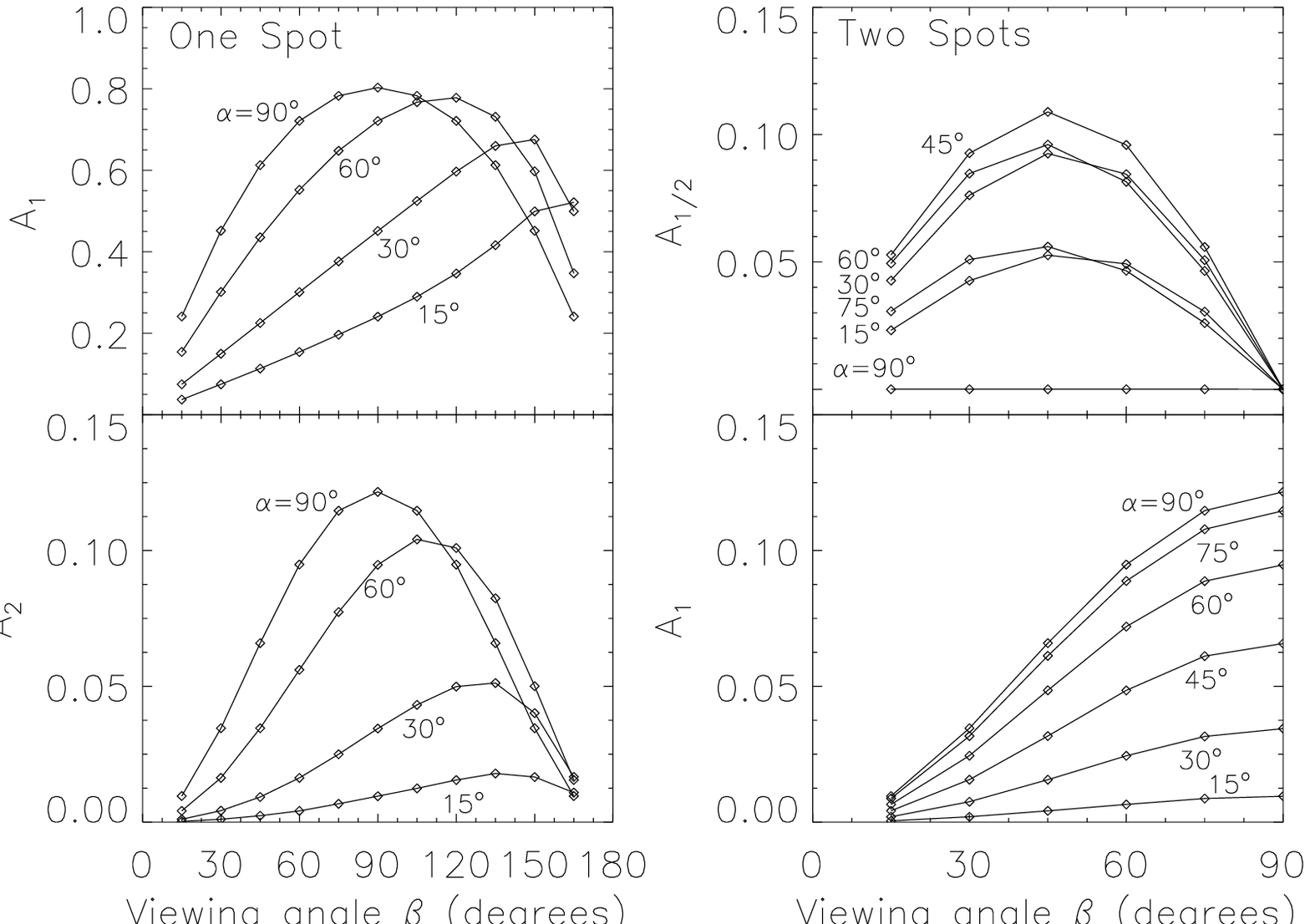,width=0.8\linewidth}}
\caption{
The predicted amplitudes of oscillations and their harmonics from 
one ({\it left panels}) and two ({\it right panels}) hot regions,
located at various angles $\alpha$ and viewed from lines of sight 
$15^\circ < \beta < 165^\circ$. Here, $\rho = 60^\circ$ and $\Omega = 300$~Hz. 
For the case of one bright region, the amplitude of both the fundamental ($A_1$) and the 
harmonic ($A_2$) decrease when $\alpha$ or $\beta$ moves away from the 
equator ($90^\circ$). For two regions, power is transferred from $A_1$ 
at $2\times\nu_{\rm spin}$ to $A_{1/2}$
at $\nu_{\rm spin}$ with decreasing $\alpha$ or $\beta$.
}
\label{fig:angle}
\end{figure*}
We use $T_{\rm col} = 2.3$~keV (as observed at infinity) in all of the
following simulations.

As the emission is peaked more strongly about the normal to the surface, 
the amplitude of the oscillation and its harmonics can be increased 
arbitrarily \citep{wml01,oze02}. The amplitudes at small compactness ($p<2$)
are particularly sensitive to the assumed beaming \citep[][]{oze02}.
For instance, if the emission is assumed to be isotropic and the star is not 
spinning, the harmonic signal disappears near $p=2.0$ \citep{ml98}, but 
reappears for $p<1.75$ when the spot is strongly lensed by the neutron star 
\citep[compare Figure~\ref{fig:compact} and][]{pfc83}. The difference is less 
stark when the star is spinning; only the minimum at $p=1.75$ is evident, and
the harmonic does not disappear (not shown). Here, we restrict ourselves to 
beaming that is described by the Hopf function, which is 
probably most relevant for these scattering-dominated atmospheres 
\citep[compare][]{mad91}. The Hopf function is slightly more 
radially peaked than isotropic emission.

We now examine in detail the effects of changing the parameters
directly related to the geometry of the hot regions, $\alpha$, $\beta$,
and $\rho$.
We first restrict both the position of the emission region 
and the observer's line of sight to the neutron star's rotational 
equator ($\alpha = \beta = 90^\circ$). 
In Figure~\ref{fig:rho}, we plot the fractional rms amplitude of the 
oscillations and their harmonics as a function of the size of the
hot region ($\rho$) for several values of the rotational frequency
as observed at infinity. 
For a single hot region, larger areas
produce lower-amplitude oscillations at the fundamental frequency ($A_1$), 
since the hot region is visible to the observer for a greater fraction of the 
rotational period. In contrast, the signal at the harmonic ($A_2$) disappears
when the bright region covers exactly half of the neutron star, and then 
increases again as $\rho$ exceeds $90^\circ$.

For two identical, antipodal emitting regions with 
$\alpha = \beta = 90^\circ$, 
no signal is seen at the neutron star's spin frequency because of 
symmetry. 
Therefore, for two bright regions we take $A_1$ to be the amplitude
of the strongest observed signal (consistent with the definition in 
Figure~\ref{fig:ulharm}), which occurs at twice the spin frequency of the 
neutron star \cite[see also][]{mil99}. For the particular case of 
$\alpha = 90^\circ$ or $\beta = 90^\circ$, 
the amplitude at $2\times\nu_{\rm spin}$ for two bright regions
with size $\rho$ is exactly the same as that at $2\times\nu_{\rm spin}$ 
for one bright region of the same size (see also 
Figure~\ref{fig:angle}). The lower right panel of Figure~\ref{fig:rho} 
demonstrates that the signal at four times the spin frequency ($A_2$) is 
always less than 5\% rms amplitude. Since the rotational frequencies
of the neutron stars we are studying must be of order 300~Hz or less if
two hot regions are responsible for the oscillations, we 
would only expect to see harmonics with $A_2 < 2\%$ amplitude, or 6\% of the 
amplitude of the main signal ($A_1$). This is similar to our best upper limits
on the amplitude of such a signal in Figure~\ref{fig:ulharm}, indicating
that we do not have the sensitivity to detect a signal at $A_2$ from 
antipodal bright regions. Therefore, we do not consider this signal further.

In Figure~\ref{fig:angle}, we consider the effect of changing the 
position of the hot region ($\alpha$) and the viewing angle ($\beta$) on 
\begin{inlinefigure}
%\epsscale{0.75}
%\plotone{f6.eps}
\centerline{\epsfig{file=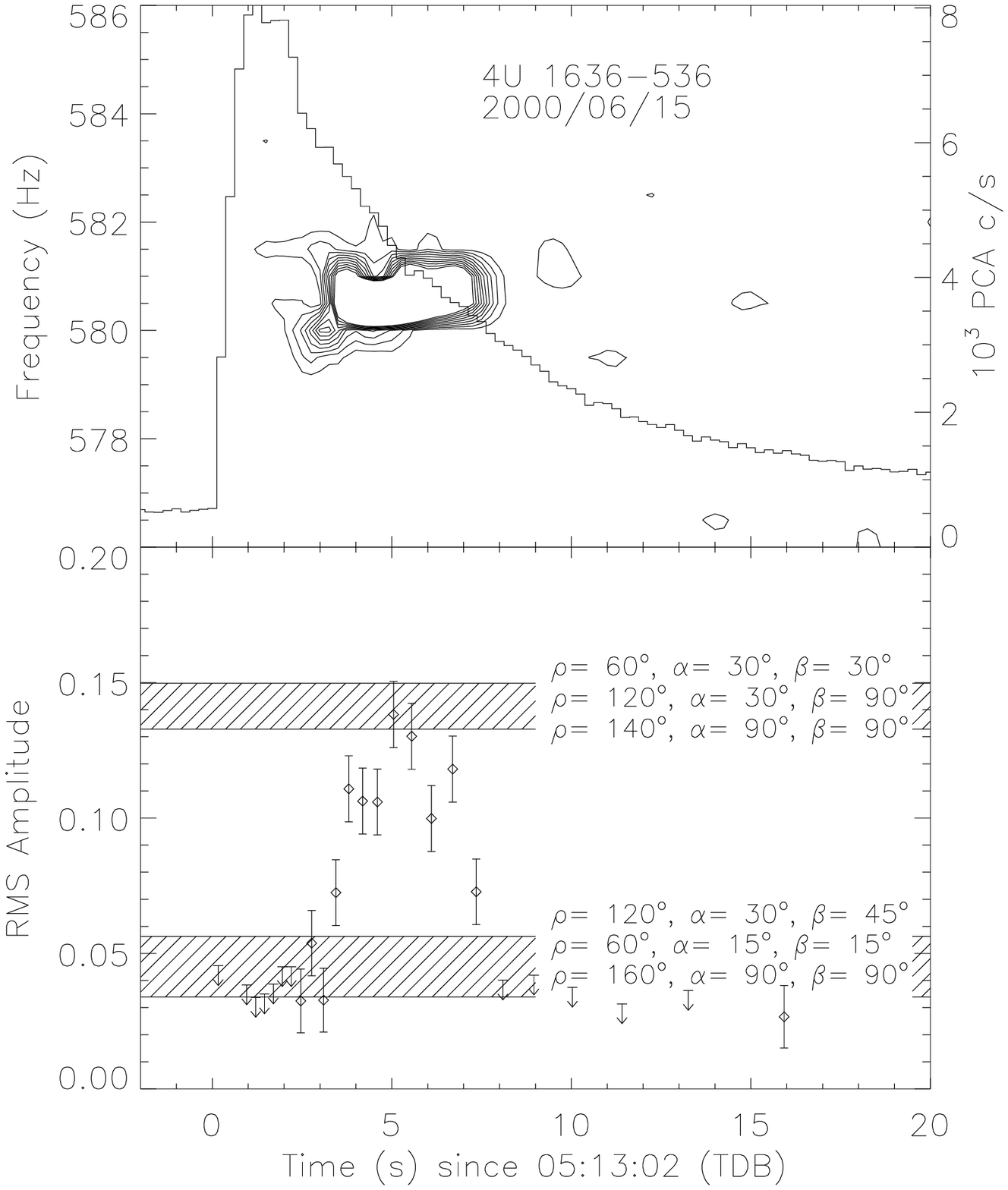,width=0.9\linewidth}}
\caption{
Same as Figure~1, for a burst from \sixb. We have indicated the range of 
$\rho$, $\alpha$, and $\beta$ that are consistent with the amplitudes 
indicated. Amplitude variations during the burst can be explained
by changes in the location or size of the bright region, but there are 
clearly degeneracies in these parameters that 
prevent us from determining them from the amplitudes alone. 
}
\label{fig:ampdeg}
\end{inlinefigure}

\noindent
the amplitude of the oscillations
and their harmonics. We display the amplitudes of signals from one or two 
bright regions with fixed size $\rho = 60^\circ$. For one region, 
the amplitude of the oscillations at both the fundamental ($A_1$)
and harmonic ($A_2$) decrease as either the observer's line-of-sight or the 
center of the hot region is moved away from the rotational equator. 
Note also that the amplitude of the harmonic decreases more
quickly with decreasing angles than that of the fundamental.
However, for $\alpha < 90^\circ$, the largest amplitude oscillation occurs 
when the observer is near the opposite pole, i.e., $\beta = 180^\circ -
\alpha$. %The decrease in the harmonic amplitude for 
%$\beta > 180^\circ - \alpha$ occurs
%only in a curved space-time; in a flat space-time, the pulse becomes
%sharper until the anisotropy disappears from view with changing $\beta$.

For two regions, one observes a signal at the spin frequency
of the neutron star whenever both the emitting regions and the observer's 
line-of-sight are away from the equator ($\alpha < 90^\circ$, 
$\beta \neq 90^\circ$ in Figure~\ref{fig:angle}). We refer to this signal 
as $A_{1/2}$, consistent with the assumption that several observed 
signals may occur at $2\times\nu_{\rm spin}$ from two antipodal hot spots 
\citep{mil99}. However, it is clear from Figure~\ref{fig:angle} that as the 
viewing angle decreases ($\beta < 90^\circ$), the amplitude of 
the signal at twice the spin frequency ($A_1$ here) decreases monotonically,
while that at the spin frequency ($A_{1/2}$) reaches a maximum at 
$\beta = 45^\circ$. In fact, $A_{1/2}$ is
larger than $A_1$ for most combinations of $\alpha$ and $\beta$.

\section{Discussion\label{sec:res}}

In this section, we compare the observed properties of the burst
oscillations with the model profiles presented in Section~\ref{sec:mod}. We 
first determine the range of
sizes and locations of the hot regions that are consistent 
with the evolution of the amplitude of the oscillations as a function of 
time (Section~4.1). We then examine the constraints 
that the lack 
of harmonic signals place on the emission geometry (Section~4.2). 

\subsection{Amplitude Changes in the Oscillations}

Figure~\ref{fig:ampdeg} shows the amplitude evolution of 
an oscillation from \sixb. Superimposed are theoretical 
amplitudes expected from several combinations of the size ($\rho$)
and location ($\alpha$) of the hot region, and of the viewing angle ($\beta$). 

It is evident that several different
combinations of these parameters can produce identical fractional amplitudes,
because a smaller hot region located near the pole of the neutron star will
on average cover the same fraction of the observer's view of the neutron 
star as will a larger region on the equator, which is obscured as it 
passes behind the neutron star (Figures~\ref{fig:rho}, \ref{fig:angle},
and \ref{fig:ampdeg}). Moreover, it is also possible
to produce smaller oscillations with small hot regions if the rest of
the neutron star is emitting flux within the PCA bandpass. Thus,
the amplitudes of the oscillations do not provide interesting 
constraints on the parameters of a brightness asymmetry, unless the amplitudes
are very large \citep{ml98,nss02}.

Changes in the amplitude of the oscillation
throughout the burst can be caused by variations in the size ($\rho$) or
position ($\alpha$) of the
bright region, and by changes in the magnitude of the brightness contrast. 
Oddly enough, the amplitude variations 
are not correlated with any particular feature of the burst
(Figures~\ref{fig:ampev} and \ref{fig:ampdeg}). If the 
oscillations are due to brightness patterns on the neutron star 
surface, then the properties of the asymmetry
must vary independent of the amount of flux emitted from the photosphere.

We also computed the mean PCA flux we would expect from the same range
of $\alpha$, $\beta$, $\rho$. The predicted count rates are 
consistent
with those observed, given $T_{\rm col} \approx 2-3$~ keV, but neglecting 
scattering effects that can greatly modify the spectrum \citep{lth84,mad91}. 
Thus, thermal emission 
from hot regions on a neutron star can explain both the mean fluxes during 
thermonuclear X-ray bursts and the amplitudes of the 
oscillations.

\subsection{Harmonic Structure of the Oscillation Profiles}

The lack of harmonic structure in the oscillation profiles 
(Figure~\ref{fig:ulharm} and Table~\ref{tab:harm}) provides
interesting and stringent constraints on the emission geometry. 
Since the amplitudes of the oscillations 
vary significantly (Figures~\ref{fig:ampev} and
\ref{fig:ampdeg}), the best means of using the harmonic amplitudes 
as a constraint is to examine
the ratios of the upper limits on their amplitudes ($A_n$) to 
that of the largest signal ($A_1$; Figure~\ref{fig:ulharm}). 

For the case of a single circular bright region, the largest harmonic signal 
occurs at twice the spin frequency of the neutron star. There are 
three ways to suppress such a signal by varying the parameters of the hot region, 
as we show below: requiring $\rho \approx 90^\circ$, 
$\alpha \lesssim 20^\circ$, or $\beta \lesssim 30^\circ$.
Results are displayed for $p=2.5$, but are not greatly sensitive to the
compactness. For instance, assuming $p=2.0$ decreases the relative amplitude
of the harmonic signals by about a factor of 1.4 (Figure~3).

In Figure~\ref{fig:harmsize}, we plot for a few values of the 
spin frequency ($\Omega$) the ratios of the amplitudes at the first harmonic 
to that at the
spin frequency ($A_2/A_1$) as a function of the 
\begin{inlinefigure}
%\epsscale{1.0}
%\plotone{f7.eps}
\centerline{\epsfig{file=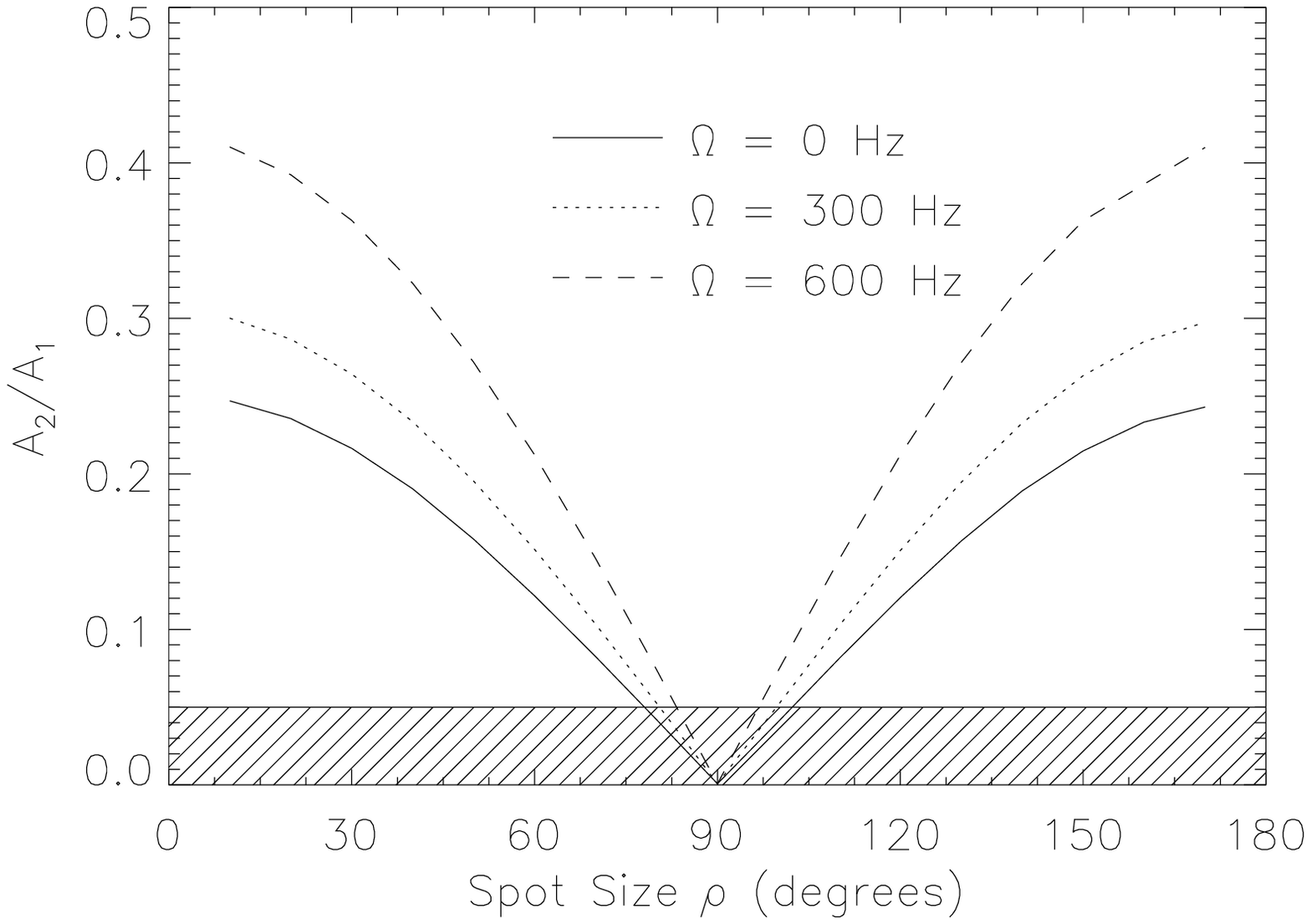,width=\linewidth}}
\caption{
The ratio of the amplitude of the harmonic signal to that of the 
fundamental, as a function of size ($\rho$) and spin period 
($\Omega$), for the one bright region at 
$\alpha = \beta = 90^\circ$. The hatched region 
indicates those values
of $A_1/A_2$ consistent with the upper limits from \slowb\ in
Figure~2, and demonstrates that a circular bright region located on
the equator must have
an angular radius of $\rho = 90^\circ \pm 10^\circ$ in order to suppress
the harmonic content of the oscillations.
}
\label{fig:harmsize}
\end{inlinefigure}

\noindent
size of the emitting region
($\rho$), assuming that the bright region and viewing angle are centered 
about the equator ($\alpha = \beta = 90^\circ$; compare Figure~\ref{fig:rho}). 
The shaded region represents the range of $A_2/A_1$ consistent with the 
upper limits from \sixb\ and \slowb\ in Figure~\ref{fig:ulharm}.
Clearly, any single bright region at $\alpha = \beta = 90^\circ$ must 
cover almost exactly half the neutron star 
($80^\circ \lesssim \rho \lesssim 100^\circ$).

As indicated in the top panel of Figure~\ref{fig:angle}, $A_2$ from a single bright
region also decreases relative 
to $A_1$ if the observer or the bright region are moved away from the 
equator. We have plotted $A_2/A_1$ as a function of these 
angles in the top panel of Figure~\ref{fig:harmang}, for a $\rho = 60^\circ$
hot region. The shaded region once 
again indicates the range of angles consistent with the upper limits
in Figure~\ref{fig:ulharm}. The harmonic components can be suppressed 
if the observer's line-of-sight is nearly aligned with the spin axis 
($\beta < 30^\circ$) or if the bright region is centered near the pole
($\alpha < 20^\circ$). It is unlikely that all of these systems are
viewed along their rotational axis. The eclipses from \mxbecl\ indicate
that it is in fact viewed near its orbital plane \citep{cw84}, which presumably 
is aligned with 
the rotational equator. Therefore, it appears that a single bright region 
must either ({\it i}) form near the rotational pole, or ({\it ii}) be 
symmetric on the neutron star surface, with an opening angle of
$\rho \approx 90^\circ$.

For two bright regions, the lack of a signal at the half-frequency 
($A_{1/2}$) provides the most interesting constraints
(compare Figure~\ref{fig:rho} and \ref{fig:angle}). It has already been 
demonstrated
that the lack of such a signal implies that the two bright regions must be 
nearly perfectly antipodal and have almost exactly the same brightness
\citep{wml01}. We estimate that the signal
at the spin frequency will have an amplitude less than 10\% that at 
$2\times\nu_{\rm spin}$ only if the hot regions are antipodal to within
$2^\circ$, and have no more than 2\% difference in their relative brightness.

In the bottom panel of Figure~\ref{fig:harmang}, we show the ratio
of the expected signals at the spin frequency of the neutron star to 
those at twice the spin frequency ($A_{1/2}/A_1$) as a function
of the angles $\alpha$ and $\beta$ for two circular hot 
\begin{inlinefigure}
%\epsscale{0.75}
%\plotone{f8.eps}
\centerline{\epsfig{file=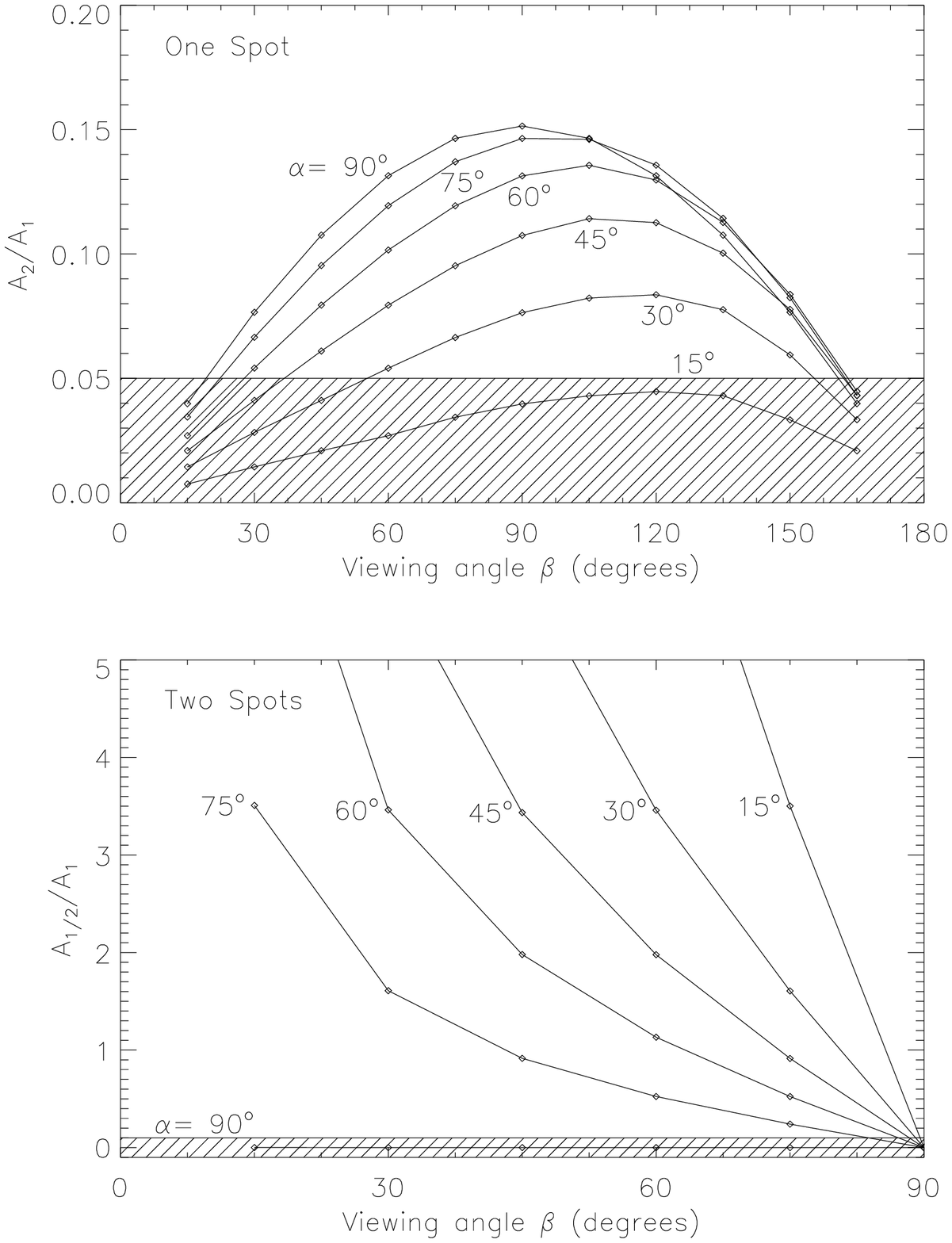,width=\linewidth}}
\caption{
The ratio of harmonic signals to the main signals as a function of the
viewing angle ($\beta$) and the location ($\alpha$) of the bright region.
Here, $\rho = 60^\circ$ and $\Omega = 300$~Hz. 
The hatched regions indicate the ranges of angles consistent with the upper 
limits in Figure~2.
{\it Top Panel}: For one bright region, the harmonic content is 
suppressed if the bright region is located at the pole ($\alpha < 20^\circ$)
or the oscillation is viewed from along the pole ($\beta < 30^\circ$). 
{\it Bottom Panel}: For two bright regions, one would observe signals at both
$\nu_{\rm spin}$ and $2\times\nu_{\rm spin}$ unless either the bright regions
are centered about the equator ($\alpha = 90^\circ$) or the 
observer's line of sight is along the equator ($\beta = 90^\circ$).
}
\label{fig:harmang}
\end{inlinefigure}

\noindent
regions.
Assuming that the strongest observed signal occurs at $2\times\nu_{\rm spin}$,
we have indicated with a shaded region those angles for which the relative
amplitudes of the $A_{1/2}/A_1$ signals are consistent with the upper limits 
in Figure~\ref{fig:ulharm}. This shows that if two bright regions are 
present, either ({\em i})
the observer's line of sight must be within a few degrees of the equator
($\beta = 90^\circ$) or ({\em ii}) 
the hot regions must be centered on the equator ($\alpha = 90^\circ$), 
in order for the two regions to appear symmetric to the observer. Regarding
the first possibility, if this model is to be applied to all of the 
sources with oscillations near 600~Hz 
\citep[][see also Table~\ref{tab:harm}]{mlp98,mil99}, it seems unlikely that 4 
of the 6 systems in Figure~\ref{fig:ulharm}
 are observed along the rotational equator 
($\alpha = 90^\circ$). Although the X-ray eclipses 
from \mxbecl\ suggest it is observed nearly edge-on, there is 
no reason to believe that \aqlxone, \sixb, and \ksxrb\ are.
If two hot regions are giving rise to the burst oscillations, 
the second possibility is the only likely one, i.e. some mechanism must
force them to form on the rotational equator. 

In addition, we can also examine the possibility that all of the 
sources form two bright regions on their surface during bursts, but
that we do not observe a signal at $2\times\nu_{\rm spin}$ in a subset of 
the sources.
By extending the bottom panel of Figure~\ref{fig:harmang} to examine
$A_{1/2}/A_1 > 15$ (not shown), we find that the observer sees only one 
of the two hot regions if they are located near the poles, and
if they are viewed along the poles. This is precisely the opposite of the 
case that would allow us to see only the signal at $2\times\nu_{\rm burst}$.
Therefore, if two hot spots are present in all of these systems, 
the two hot regions must form only at the equator or near the poles.

%%%
We have not explored how shadowing and scattering by the accretion flow 
modifies the pulse profiles, because the integration 
technique that we used to compute the general relativistic photon trajectories
does not allow us to check whether a photon is intercepted along its path
toward the observer. We believe that shadowing by the accretion disk is not likely 
to affect the relative amplitudes of the harmonics. The disk would probably lie 
along the rotational equator of the neutron star, since the accreted material 
exerts significant torques on the star over the lifetime of the system. Therefore, 
the constant fraction of the emitting region that is below the equator will be 
obscured at every rotational phase, which will suppress all of the harmonics equally 
\citep[e.g.,][]{ss01}. If the disk is for some reason mis-aligned with the 
rotational equator, the disk would occult the bright region. This would tend to narrow 
the pulse shape, increasing the relative amplitude of the harmonics. Thus, we find no 
reason to expect that shadowing by the accretion disk would serve to suppress the 
harmonic content of the oscillations, although more detailed simulations are warranted 
to test our hypotheses.

On the other hand, scattering in an accretion disk \citep{ss01} or in a spherical 
corona around the neutron star \citep[e.g.][]{bl87,bus88,mil00} could explain the 
lack of harmonic content in the oscillations. The scattering both damps the peak 
amplitude of the signal from the bright regions, and allows the observer to 
receive flux from the bright spots over a larger fraction of the neutron star's 
rotational period. As a result, the fundamental and odd harmonics are suppressed, 
and the even harmonics are nearly completely removed from the pulse profile. Although 
these scattering effects can not be unambiguously 
identified in the current study, they could be revealed by future measurements of 
changes in the phase and profile of the oscillations as a function of energy.

\section{Conclusions and Further Implications}

We examined the amplitudes and profiles of a sample of 59 burst
oscillations observed from 6 different neutron star LMXBs with the PCA
aboard \rxte. We focus on the long oscillation trains observed during the 
declines of bursts. We find that these oscillations have rms amplitudes as high
as 15\%, but that by 15~s into the burst, the oscillations drop suddenly below our 
significance threshold. Other than this general trend, variations as large as a 
factor of two in the fractional amplitudes of the oscillations do not correspond to 
similar changes in the underlying flux from the bursts (Figure~\ref{fig:ampev}). 

We computed pulse profiles for each oscillation. On average, the rms 
amplitudes of the oscillations are 2--10\%, and the
typical folded profile contains $7\times10^4$ photons. The most
significant individual oscillation has a Leahy-normalized power of 615 
in $1\times10^5$ photons, for an amplitude of 8\%. However, the eclipsing 
source \mxbecl\ routinely produces 10\% oscillations in weak 
bursts ($1\times10^4$ photons). 
We also produced summed pulse profiles for each source. Those from 
\sixb\ and \slowb\ contained more than $1\times10^6$ photons, which 
allowed us to place upper limits on the 
amplitudes of harmonic and half-frequency signals of less than 0.3\% and 
0.6\% respectively (95\% confidence). These upper limits are 
6\% and 10\% of the amplitudes of the strongest signals 
(Figure~\ref{fig:ulharm} and Table~\ref{tab:harm}). Thus, the profiles of the
oscillations are remarkably sinusoidal.

We then derived theoretical lightcurves of pulsations from one
or two circular bright regions on the surface of a rapidly rotating
star. Comparing the observed and theoretical light curves, we find
that the lack of harmonic content in the oscillations can be explained
for a single bright region if it either lies near the rotational pole
(Figure~\ref{fig:harmang}, {\it top}) or covers nearly half the neutron star
(Figure~\ref{fig:harmsize}). This result is fairly 
insensitive to the compactness of the neutron stars (Figure~\ref{fig:compact}).

If two antipodal hot regions give rise to the flux oscillations, 
the situation is even more restricted 
(Figure~\ref{fig:harmang}, {\it bottom}). The bright
regions would have to be located either ({\it i}) near the poles such that
only the signal at $\nu_{\rm spin}$ is visible, or ({\it ii}) on the equator,
so that only the signal at $2\times\nu_{\rm spin}$ is visible.
A mechanism would have to be invoked that prevents bright regions from
being formed at intermediate latitudes. Furthermore, the flux
difference between the two hot regions would need to be less than
$2\%$ to be consistent with the lack of harmonic signals.

The geometric constraints implied by the sinusoidal pulse shapes
present a challenge for theoretical models for producing brightness
patterns on the neutron star's surface. Models that invoke 
uneven heating or cooling \citep{str97} or hydrodynamical 
instabilities in a geostrophic flow \citep{slu01} do not propose 
natural mechanisms for constraining the size and location of 
brightness asymmetries (Figures~\ref{fig:harmsize} and
\ref{fig:harmang}).

The most promising mechanism for producing symmetric anisotropies with
restricted geometries is the excitation of global modes that propagate 
in the neutron star ocean, with an azimuthal dependence of the form 
$e^{im\phi}$ (e.g., Heyl 2002). The density fluctuations of an $m=1$ 
oscillation would divide the neutron star into symmetric halves, while
an $m=2$ mode would
naturally form on the equator, its symmetry ensured by the Coriolis
forces at higher latitudes. However, no physical mechanism to convert
density fluctuations on the stellar surface to flux oscillations has
been proposed. Furthermore, the surface velocities of the global
excitations, and hence their observed frequencies, depend sensitively on 
the vertical structure of the neutron star atmosphere 
\citep[e.g.,][]{mt87,bc95,sl96}. Therefore, it is important to model the outer
layers of a neutron star during thermonuclear burning and subsequent
cooling in order to establish whether the frequencies of these modes
are similar to those required to explain the frequency drift of the
burst oscillations.

On the other hand, the harmonic content of the pulsations can also be suppressed
if the signal from the neutron star scatters in the accretion flow before it
reaches the observer \citep{ss01, bl87, bus88, mil00}. Further signatures of 
scattering should be sought in the energy dependence of the profiles of burst
oscillations.

\acknowledgements
We thank Cole Miller for sharing the results of his simulations to compare 
with our work, Dimitrios Psaltis for his advice and encouragement while we 
pursued this project, and the referee for insightful comments and suggestions. 
Derek Fox, Duncan Galloway, and Pavlin Savov made 
important contributions to the data analysis underlying this work. 
We were supported in part by NASA, under contract 
NAS 5-30612 and grant NAG 5-9184.

\begin{deluxetable}{lcccccccc}
\tabletypesize{\small}
\tablecolumns{9}
\tablewidth{0pc}
\tablecaption{Harmonic Amplitudes of Burst Oscillations\label{tab:harm}}
\tablehead{
\colhead{(1)} & \colhead{(2)} & \colhead{(3)} & \colhead{(4)} & 
\colhead{(5)} & \colhead{(6)} & 
\colhead{(7)} & \colhead{(8)} & \colhead{(9)} \\
\colhead{Source} & \colhead{$\nu_1$} & \colhead{No. Osc.} & \colhead{Counts} & 
\colhead{Background} & \colhead{$A_{1/2}$} & 
\colhead{$A_1$} & \colhead{$A_{3/2}$} & \colhead{$A_2$} 
}
\startdata
\sixb & 581 & 17\tablenotemark{a} & $1.1\times10^6$ & $1.3\times10^5$ & $< 0.6$ & 5.4(3) & $< 0.5$ & $< 0.3$ \\
\mxbecl & 567 & 3 & $2.8\times10^4$ & $6.1\times10^3$ & $< 2.7$ & 9.3(8) & $< 2.8$ & $< 2.7$ \\
\aqlxone & 549 & 3 & $4.2\times10^5$ & $2.0\times10^4$ & $< 0.6$ & 3.3(1) & $< 0.5$ & $< 0.5$ \\
\ksxrb & 524 & 4 & $2.5\times10^5$ & $4.5\times10^4$ & $< 1.2$ & 4.7(2) & $< 0.9$ & $< 0.6$ \\ 
\slowb & 363 & 24\tablenotemark{b} & $1.6\times10^6$ & $2.3\times10^5$ & $< 0.6$ & 5.5(1) & $< 0.6$ & $< 0.3$ \\
\sevenb & 329 & 8 & $6.1\times10^5$ & $5.6\times10^4$ & $< 0.6$ & 4.6(2) & $< 0.7$ & $< 0.7$ \\
\enddata
\tablecomments{Columns are as follows: (1) Source name. (2) The approximate 
frequency of the observed oscillations. (3) Number of bursts with oscillations
used to make a combined profile. (4) Total number of counts in the profile, 
including background. (5) Estimated background counts in the profile. 
(6-9) Percent fractional rms amplitudes,
or 95\% upper limits on amplitudes at $n=$0.5,1,1.5, and 2 times the main
frequency ($\nu_1$).}
\tablenotetext{a}{11 oscillations were used to constrain $A_{1/2}$ and 
$A_{3/2}$, for a total of $9.8\times10^5$ counts with $1.0\times10^5$ 
counts background.}
\tablenotetext{b}{13 oscillations were used to constrain $A_{1/2}$ and 
$A_{3/2}$, for a total of $1.2\times10^6$ counts with $1.9\times10^5$ counts
background.}
\end{deluxetable}


\begin{thebibliography}{0}
\bibitem[Alpar et al.(1982)]{alp82} Alpar, M. A., Cheng, A. F., Ruderman,
	M. A., \& Shaham, J. 1982, \nat, 300, 728
\bibitem[Bildsten(1995)]{bil95} Bildsten, L. 1995, \apj, 438, 852
\bibitem[Bildsten(1998)]{bil98} Bildsten, L. 1998, \apj, 501, L89
\bibitem[Bildsten \& Cutler(1995)]{bc95} Bildsten, L. \& Cutler, C. 1995, \apj, 
	449, 800
\bibitem[Brainerd \& Lamb(1987)]{bl87} Brainerd, J. \& Lamb, F. K. 1987, \apj, 
	317, L33
\bibitem[Braje et al.(2000)]{brr00} Braje, T. M., Romani, R. W., \&
	Rauch, K. P. 2000, \apj, 531, 447
\bibitem[Bussard et al.(1988)]{bus88} Bussard, R. W., Weisskopf, M. C., Elsner, 
	R. F., \& Shibazaki, N. 1988, \apj, 327, 284
\bibitem[Chandrasekhar(1960)]{cha60} Chandrasekhar, S. 1960, 
	{\it Radiative Transfer} (Dover)
\bibitem[Cominsky \& Wood(1984)]{cw84} Cominsky, L. R. \& Wood, K. S. 1984,
	\apj, 283, 765
\bibitem[Cumming \& Bildsten(2000)]{cb00} Cumming, A. \& Bildsten, L. 2000, 
	\apj, 544, 453
\bibitem[Cumming et al.(2002)]{cum02} Cumming, A., Morsink, S. M., Bildsten,
	L., Friedman, J. L., \& Holz, D. E. 2002, \apj, 564, 343
%\bibitem[Deeter, Pravdo, \& Boynton(1981)]{dbp81} Deeter, J. E., Pravdo, S. H.,
%    Boynton, P. E. 1981, \apj, 247, 1003
%\bibitem[Ford(1999)]{for99} Ford, E. C. 1999, \apj, 519, L73
%\bibitem[Galloway et al.(2001)]{gal01} Galloway, D. K., Chakrabarty, D.,
%	Muno, M. P., \& Savov, P. 2001, \apj, 549, L85
\bibitem[Giles et al.(2002)]{gil02} Giles, A. B., Hill, K. M., Strohmayer, T. 
	E., \& Cummings, N. 2002, \apj, 568, 279
\bibitem[Heyl(2002)]{hey02} Heyl, J. S. 2002, \mnras, submitted, 
	astro-ph/0108450
\bibitem[Jahoda et al.(1996)]{jah96} Jahoda, K., Swank, J. H., Giles, A. B., 
	Stark, M. J., Strohmayer, T., Zhang, W., \& Morgan, E. H. 1996, 
	SPIE, 2808, 59
\bibitem[Kuulkers et al.(2002)]{kul02} Kuulkers, E., Homan, J., van der Klis,
	M., Lewin, W. H. G., \& M\'{e}ndez, M. 2002, \aap, 382, 947
\bibitem[Leahy \etal(1983)]{lea83} Leahy, D. A., Darbro, W., Elsner, R. F.,
        Weisskopf, M. C., Sutherland, P. G., Kahn, S., \& Grindlay, J. E.
        1983, \apj, 266, 160
\bibitem[Lewin et al.(1993)]{lvt93} Lewin, W. H. G., van 
	Paradijs, J., \& Taam, R. E. 1993, Space Sci. Rev., 62, 223
\bibitem[London et al.(1984)]{lth84} London, R. A., Taam, R. E.,
	\& Howard, W. M. 1984, \apj, 287, L27
\bibitem[Madej(1991)]{mad91} Madej, J. 1991, \apj, 376, 161
\bibitem[Manchester \& Taylor(1977)]{mt77} Manchester, R. N., \& Taylor,
    J. H. 1977, Pulsars(San Francisco: W. H. Freeman and Co.)
\bibitem[McDermott \& Taam(1987)]{mt87} McDermott, P. N., \& Taam, R. E. 1987,
	\apj, 318, 278	
\bibitem[Miller(1999)]{mil99} Miller, M. C. 1999, \apj, 515, L77
\bibitem[Miller(2000)]{mil00} Miller, M. C. 2000, \apj, 537, 342
\bibitem[Miller \& Lamb(1998)]{ml98} Miller, M. C. \& Lamb, F. K. 1998, 
	\apj, 499, L37
\bibitem[Miller et al.(1998)]{mlp98} Miller, M. C., Lamb, F. K.,
	\& Psaltis, D. 1998, \apj, 508, 791
\bibitem[Muno et al.(2001)]{mun01} Muno, M. P., Chakrabarty, D., Galloway,
	D. K., \& Savov, P. 2001, \apj, 553, L157
\bibitem[Muno et al.(2000)]{mun00} Muno, M. P., Fox, D. W., Morgan, E. H., 
	\& Bildsten, L. 2000, \apj, 542, 1016
\bibitem[Muno et al.(2002)]{mgc02} Muno, M. P., Galloway, D. K. Chakrabarty,
	D., \& Psaltis, D. 2002, \apj, to appear in 580, No. 2, astro-ph/0204320
\bibitem[Nath et al.(2002)]{nss02} Nath, N. R., Strohmayer, T. E., \& 
	Swank, J. H. 2002, \apj, 564, 353
\bibitem[\"{O}zel(2002)]{oze02} \"{O}zel, F. 2002, \apj, in press, 
	astro-ph/021158
%\bibitem[\"{O}zel, Psaltis, \& Kaspi(2001)]{opk01} \"{O}zel, F., 
%	Psaltis, D., \& Kaspi, V. M. 2001, \apj\, 563, 255
\bibitem[Pechenick et al.(1983)]{pfc83} Pechenick, K. R., Ftaclas, C., \&
	Cohen, J. M. 1983, \apj, 274, 846
\bibitem[Press \etal\ (1992)]{pre92} Press, W. H., Teukolsky, S. A.,
        Vetterling, W. T. \& Flannery, B. P. 1992, Numerical Recipes
        in C, 2nd Ed. (Cambridge: Cambridge University Press)
\bibitem[Radhakrishnan \& Srinivasan(1982)]{rs82} Radhakrishnan, V.
         \& Srinivasan, G. 1982, Curr. Sci., 51, 1096
\bibitem[Sazonov \& Sunyaev(2001)]{ss01} Sazonov, S. Y. \& Sunyaev, R. A. 2001, \aap,
	373, 241
\bibitem[Schoelkopf \& Kelley(1991)]{sk91} Schoelkopf, R. J. \& 
	Kelley, R. L. 1991, \apj, 375, 696
\bibitem[Spitkovsky et al.(2002)]{slu01} Spitkovsky, A., 
	Levin, Y., \& Ushomirsky, G. 2002, \apj, 566, 1018
\bibitem[Standish et al.(1992)]{sta92} Standish, E.~M., Newhall, X.~X., 
	Williams, J.~G., \& Yeomans, D.~K.
	1992, in Explanatory Supplement to the Astronomical Almanac, ed. P.~K.
	Seidelmann (Mill Valley: University Science), 279
%\bibitem[Strohmayer(1999)]{str99} Strohmayer, T. E. 1999, \apj, 523, L51
\bibitem[Strohmayer(2001)]{str01} Strohmayer, T. E. 2001,
	Adv. Space. Res., 28, 511
\bibitem[Strohmayer et al.(1997a)]{str97} Strohmayer, T. E., Jahoda, K., 
	Giles, A. B., \& Lee, U. 1997, \apj, 486, 355
\bibitem[Strohmayer \& Lee(1996)]{sl96} Strohmayer, T. E. \& Lee, U. 1996, \apj,
	467, 773
\bibitem[Strohmayer \& Markwardt(1999)]{sm99} Strohmayer, T. E. \& 
	Markwardt, C. B. 1999, \apj, 516, L81
\bibitem[Strohmayer \& Markwardt(2002)]{sm02} Strohmayer, T. E. \&
	Markwardt, C. B. 2002, \apj, submitted
\bibitem[Strohmayer et al.(1997b)]{szs97} Strohmayer, T. E., Zhang, 
	W., \& Swank, J. H. 1997, \apj, 487, L77
%\bibitem[Strohmayer et al.(1998)]{str98} Strohmayer, T. E., Zhang, 
%	W., Swank, J. H. \& Lapidus, I. 1998a, \apj, 503, L147
\bibitem[Strohmayer et al.(1996)]{str96} Strohmayer, T. E., Zhang, W., Swank, 
	J. H., Smale, A., Titarchuk, L., Day, C., \& Lee, U. 1996, \apj, 
	469, L9
\bibitem[van der Klis(2000)]{vdk00} van der Klis, M. 2000, \araa, 38, 717
\bibitem[Vaughan et al.(1994)]{vau94} Vaughan, B. A. et al. 1994, \apj, 
	435, 362
\bibitem[Weinberg et al.(2001)]{wml01} Weinberg, N., Miller, M. C., \&
 	Lamb, D. Q. 2001, \apj, 546, 1098
\bibitem[White \& Zhang(1997)]{wz97} White, N.~E. \& Zhang, W. 1997,
	\apj, 490, L87
\end{thebibliography}
\end{document}